\documentclass[a4paper,11pt]{article}
\pdfoutput=1 

\usepackage{jcap/jcappub} 
\usepackage{graphicx}
\usepackage{color}
\usepackage{hyperref}
\usepackage{dcolumn}
\usepackage{multirow}
\usepackage{caption}
\usepackage{subcaption}
\usepackage{bbold}

\newcommand{\be}{\begin{equation}}
\newcommand{\ee}{\end{equation}}
\newcommand{\bea}{\begin{eqnarray}}
\newcommand{\eea}{\end{eqnarray}}
\newcommand{\bit}{\begin{itemize}}
\newcommand{\eit}{\end{itemize}}
\newcommand{\bfi}{\begin{figure}}
\newcommand{\efi}{\end{figure}}
\newcommand{\bfic}{\begin{figure*}}
\newcommand{\efic}{\end{figure*}}
\newcommand{\bce}{\begin{center}}
\newcommand{\ece}{\end{center}}
\newcommand{\bt}{\begin{table}}
\newcommand{\et}{\end{table}}
\newcommand{\btb}{\begin{tabular}}
\newcommand{\etb}{\end{tabular}}
\newcommand{\bwt}{\begin{widetext}}
\newcommand{\ewt}{\end{widetext}}

\newcommand{\UD}[2]{\ensuremath{^{#1}_{\phantom{#1} #2}}}
\newcommand{\DU}[2]{\ensuremath{_{#1}^{\phantom{#1} #2}}}

\newcommand{\bi}[1]{\ensuremath{\textbf{#1}}}
\newcommand{\bb}[1]{\ensuremath{\mathbf{#1}}}
\newcommand{\curl}{\ensuremath{\textrm{curl }}}

\newcolumntype{d}[1]{D{.}{\cdot}{#1} }

\title{ On the vacuum Einstein equations along curves with a discrete local rotational and reflection symmetry }
\author[a]{Miko\l{}aj Korzy\'nski,}
\affiliation[a]{
Center for Theoretical Physics \\
Polish Academy of Sciences \\
Al. Lotnik\'o{}w 32/46, 02-668 Warsaw \\
Poland}
\emailAdd{korzynski@cft.edu.pl}

\author[b]{Ian Hinder}
\affiliation[b]{
Max-Planck-Institut f\"ur Gravitationsphysik\\
Albert-Einstein-Institut \\ 
Am M\"uhlenberg 1, D-14476 Golm \\
Germany}
\emailAdd{ian.hinder@aei.mpg.de}

\author[c,d]{and Eloisa Bentivegna}
\affiliation[c]{
Dipartimento di Fisica e Astronomia\\
Universit{\`a} degli Studi di Catania\\
Via S.~Sofia 64, 95123 Catania\\
Italy}
\affiliation[d]{
INFN\\
Sezione di Catania\\
Via S.~Sofia 64, 95123 Catania\\
Italy}
\emailAdd{eloisa.bentivegna@ct.infn.it}

\abstract{

We discuss the possibility of a dimensional reduction of the Einstein equations in $S^3$ 
black-hole lattices.  It was reported in previous literature that the evolution of spaces containing curves
of local, discrete rotational and reflection Symmetry (LDRRS) can be carried out via a system of ODEs along these curves.  However, 3+1
Numerical Relativity computations demonstrate that this is not the case, and we show analytically that this is due to the
presence of a tensorial quantity which is not suppressed by the symmetry. 
We calculate the term analytically, and verify numerically for an 8-black-hole lattice that it
fully accounts for the anomalous results,
and thus quantify its magnitude in this specific case.
The presence of this term prevents the exact evolution of these spaces via previously-reported methods which do not
involve a full 3+1 integration of Einstein's equation.
}

\begin{document}

\maketitle
\flushbottom

\section{$S^3$ black-hole lattices and dimensional reduction} 
Black hole lattices, and in particular those conformally related to the 3-sphere $S^3$,
have recently been the object of several studies~\cite{Clifton:2012qh,Bentivegna:2012ei,
Clifton:2013jpa,Clifton:2014lha,Clifton:2014mza,Liu:2015bwa}
serving as toy models for the evolution of inhomogeneous universes~\cite{Bentivegna:2012ei,Clifton:2013jpa}, for the propagation of
gravitational waves in periodic spaces~\cite{Clifton:2014lha}, for the exploration of cosmological models of 
non-trivial topology~\cite{Clifton:2014mza}, and as an application of Regge calculus~\cite{Liu:2015bwa}.

In~\cite{Clifton:2013jpa}, in particular, it was pointed out that the existence of a time-symmetric spatial
hypersurface in these models, in addition to the high degree of spatial symmetry of the cell
edges, implied that the evolution of the proper length of these subspaces was
governed by a system of ordinary differential equations, and was thus decoupled from the
surrounding spacetime. Here, however, we show that this is not the case: the symmetry is
not sufficient to suppress one term, proportional to the curl of the magnetic part of the
Weyl tensor, which contains spatial derivatives of the extrinsic curvature, and therefore
prevents the reduction of the evolution equations to a simple, localized ODE system.

In section~\ref{sec:lrs}, we 
present the arguments of~\cite{Clifton:2013jpa} and 
the corresponding ODE system. In section \ref{sec:num}, we integrate
the Einstein equations numerically in 3+1 dimensions, and show that the comparison 
between the result and the solution of the ODE system features an anomaly which 
converges to a non-zero value in the continuum limit. In section \ref{sec:curl},
we rederive the equations of motion of $S^3$ black-hole lattices
on the cell edges, and illustrate the origin of the curl term. We also show that in the 8-black-hole case, the additional term initially vanishes together with its first two
time derivatives, but its third time derivative does not, thus providing the final piece of evidence that the ODE system from~\cite{Clifton:2013jpa} does not hold for the 
8-black-hole lattice. Finally, we show
in section \ref{sec:numu33} that the anomaly measured numerically in section \ref{sec:num}
coincides with the curl term derived in section \ref{sec:curl}, and provide
a fitting formula for representing this term analytically, so that the ODE system can still
be used in combination with this source term.
We provide some conclusions in section \ref{sec:concl}. 

We will use the following index conventions throughout the paper: Greek indices $\mu, \, \nu, \, \dots$, 
will run from 0 to 3 and denote the spacetime objects. 
Objects defined on a spatial slice of dimension 3 will be denoted using Latin indices $i, \, j, \, \dots$.  Tensorial quantities will be 
sometimes written in the index notation ($\gamma_{ij}$, $K_{ij}$) and sometimes in the index-free notation using the bold typeface ${\boldsymbol \gamma}$, $\bb{K}$.

\section{Definition and properties of the LDRRS curves}
\label{sec:lrs}

In this section we will introduce the mathematical formalism required to formulate the result of this paper. We will first restate the definition of a curve with a 
local, discrete rotational and reflection symmetry (LDRRS), discuss the effects of these symmetries and present the ODEs derived in \cite{Clifton:2013jpa}.
The derivation of the reduced Einstein equations in \cite{Clifton:2013jpa} has been performed using the orthonormal frame approach in the version
given by van Elst and Uggla \cite{vanElst-Uggla}. 
This approach is less known than the standard ADM formalism  and certainly less useful in numerical investigations where we have direct access to the 3-metric and the extrinsic
 curvature on a time slice, rather than the Ricci rotation coefficients or the commutation relations of an orthonormal frame. We shall therefore present here the derivation of the reduced 
 evolution equations from the ADM equations.
Naturally, the final result does not depend on the formalism used.

Consider the vacuum ADM equations in the normal gauge (corresponding to Gaussian normal coordinates, shift $\beta^i=0$, lapse $\alpha=1$)
\bea
\dot \gamma_{ij} &=& -2K_{ij} \label{eq:ADM1}\\
\dot K_{ij} &=& R_{ij} - 2K_{ik}\,K\UD{k}{j} + K\,K_{ij} \label{eq:ADM2}
\eea
where $\gamma_{ij}$ is the 3-metric on a spacelike hypersurface $\Sigma$, $K_{ij}$ denotes the extrinsic 
curvature of this hypersurface, $K = K\UD{i}{i}$ its trace and $R_{ij}$ is the 3-dimensional Ricci tensor of $\gamma_{ij}$.
The constraint equations read
\bea
R + K^2 - K_{ij}\,K^{ij} &=& 0 \label{eq:constraint1}\\
\left(K^{ij} - K\,\gamma^{ij}\right)_{;i} &=& 0 \label{eq:constraint2}
\eea
where $R = R\UD{i}{i}$ and the covariant derivative is taken with respect to $\gamma_{ij}$~\cite{adm, 1979sgrr.work...83Y, Anninos:lr}.

Following \cite{Clifton:2013jpa}, we assume that on a given $\Sigma$ there exists a curve $\lambda$ and a discrete group of symmetries $G$ in the form of 
discrete $n$-fold rotations about $\lambda$ together with reflections through planes passing through $\lambda$. More precisely, we assume that for each $a \in G$
there exists a mapping $R_a$ defined on a neighbourhood of $\lambda$ which preserves both the 3-metric and the extrinsic curvature:
\bea
R_a^*\,\boldsymbol{\gamma} &=& \boldsymbol{\gamma} \\
R_a^*\,\bb{K} &=& \bb{K},
\eea
$R_a^*$ denoting the pullback of a tensor by $R_a$. We assume it leaves every point in $\lambda$ invariant:
\bea
R_a(p) = p\qquad \textrm {if } p\in \lambda.
\eea
It follows from the assumptions above that $R_a$ induces a mapping on the tangent space at every point $p \in \lambda$, i.e. $R_a^*: T_p\Sigma \mapsto T_p\Sigma$, which leaves
both $\gamma_{ij}$ and $K_{ij}$ at $p$ invariant.

We also assume that the action
of $G$ on $T_p\Sigma$ is the action of the group of discrete rotations and reflections. This assumption may be phrased in the following way: let $r$ generate the rotations and let $m$
and $r$ generate the reflections. Then, in an appropriately chosen, properly oriented orthonormal frame $\bi{e}_i$ in $T_p \Sigma$, 
we have
\bea
R_r^* (\bi{e}_1) &=& \bi{e}_1  \label{eq:R_r1} \\
R_r^* (\bi{e}_2) &=& \cos \frac{2\pi}{n}\,\bi{e}_2 +\sin \frac{2\pi}{n}\,\bi{e}_3\label{eq:R_r2}\\
R_r^* (\bi{e}_3) &=& -\sin \frac{2\pi}{n}\,\bi{e}_2 +\cos \frac{2\pi}{n}\,\bi{e}_3  \label{eq:R_r3}
\eea
and
\bea
R_m^* (\bi{e}_1) &=& \bi{e}_1 \label{eq:R_m1} \\
R_m^* (\bi{e}_2) &=& -\bi{e}_2 \label{eq:R_m2} \\
R_m^* (\bi{e}_3) &=& \bi{e}_3 \label{eq:R_m3} 
\eea
where $\bi{e}_1$ has been chosen to be tangent to $\lambda$. Note that $\bi{e}_1$, $\bi{e}_2$ and $\bi{e}_3$ are not coordinate basis vectors.

It is straightforward to see that the time development of $\lambda$ under the vacuum Einstein equations in normal coordinates will be a curve with local rotational and reflection symmetry.
In \cite{Clifton:2013jpa} the authors prove that the assumption of invariance under (\ref{eq:R_r1})--(\ref{eq:R_m3}) restricts the form of vectors and tensors at points 
lying on $\lambda$. 
In particular, a vector field $X^i$ invariant with respect to rotation (\ref{eq:R_r1})--(\ref{eq:R_r3}) has
to be aligned along the curve $\lambda$ at every point $p\in \lambda$:
\bea
\bi{X}_p = X_1 \,\bi{e}_1 \label{eq:GinvX}
\eea
and every rotation-invariant symmetric 2-tensor $S_{ij} = S_{(ij)}$ is a combination of the metric and a symmetric traceless tensor:
\bea
\bi{S}_p = \frac{S\UD{i}{i}}{3} \,\boldsymbol{\gamma} + S_{11}\left(\boldsymbol{\alpha}_1\otimes \boldsymbol{\alpha}_1-\frac{1}{2}\boldsymbol{\alpha}_2\otimes \boldsymbol{\alpha}_2
-\frac{1}{2}\boldsymbol{\alpha}_3\otimes 
\boldsymbol{\alpha}_3 \right), \label{eq:GinvS}
\eea
where $S\UD{i}{i}$ is the trace of $S$, $S_{11}$ is the $(1,1)$ component of $S_{ij}$ in the orthonormal frame $\bi{e}_i$ and $\boldsymbol\alpha^i$ is the dual co-frame of $
\bi{e}_i$, i.e. $\boldsymbol{\alpha}^i(\bi{e}_j) = \delta\UD{i}{j}$.
On the other hand, every rotation- and reflection-invariant antisymmetric 2-tensor $A_{ij} = A_{[ij]}$ has to vanish at $p$:
\bea
\bi{A}_p = 0. \label{eq:GinvA}
\eea

It follows quite easily from (\ref{eq:GinvX})  that $\lambda$ must be a geodesic with respect to $\gamma_{ij}$. Indeed, let $\bi{v}$ be a tangent vector to $\lambda$ 
in any parametrization. The vector $\nabla_{\bi{v}} \bi{v}$ is rotation-invariant since both the curve $\lambda$ and the metric $\gamma_{ij}$ are invariant as well. From (\ref{eq:GinvX})
it must be proportional to
$\bi{e}_1$, and thus also to $\bi{v}$ itself. After a suitable reparametrisation we obtain $\nabla_{\bi{v}} \bi{v} = 0$.

In~\cite{Clifton:2013jpa}, following~\cite{vanElst-Uggla}, a formalism was presented for simplifying the system
(\ref{eq:ADM1})--(\ref{eq:ADM2}) on a LDRRS curve. In this case, the only degrees
of freedom of the metric tensor not suppressed by the symmetries are:
\begin{align}
 a_\parallel &= \sqrt{\boldsymbol{\gamma}\left(\bi{Z}_1,\bi{Z}_1\right)} \label{eq:apar} \\
  a_\perp &= \sqrt{\boldsymbol{\gamma}\left(\bi{Z}_2,\bi{Z}_2\right)} = \sqrt{\boldsymbol{\gamma}\left(\bi{Z}_3,\bi{Z}_3\right)}, \label{eq:aperp}
\end{align}
where the vectors $\bi{Z}_1$, $\bi{Z}_2$ and $\bi{Z}_3$ are initially equal to $\bi{e}_1$, $\bi{e}_2$ and $\bi{e}_3$ respectively, and their components 
are assumed to be constant in time in the normal coordinate basis ($\bi{Z}_1$, $\bi{Z}_2$ and $\bi{Z}_3$ are the coordinate basis vectors if we choose
the initial coordinate system appropriately). Note that the functions $a_\parallel$ 
and $a_\perp$ are sufficient to reconstruct the metric tensor at any time:
\bea
\boldsymbol{\gamma} =  a_\perp(t)^2\left(\boldsymbol{\omega}_1\otimes\boldsymbol{\omega}_1 + \boldsymbol{\omega}_2\otimes\boldsymbol{\omega}_2\right) +  a_\parallel(t)^2\,
\boldsymbol{\omega}_3\otimes\boldsymbol{\omega}_3, 
\eea
where $\boldsymbol{\omega}_i$ is the dual co-frame of $\bi{Z}^i$, with constant components in a normal coordinate system basis. 
The functions evolve according to: 
\begin{align}
  \frac{\ddot a_\parallel}{a_\parallel} &= \frac{2}{3} E_+, \label{eq:eapardot}\\
  \frac{\ddot a_\perp}{a_\perp} &= -\frac{1}{3} E_+ . \label{eq:aperpdot}
\end{align}
$E_+$ is the only surviving component of the electric part of the Weyl tensor, given by  
\bea
&&E_+ = -\frac{3}{2} E_{ij}\,\bi{e}_1^i\,\bi{e}_1^j \label{eq:Eplusdef} \\
&&E_{\mu\nu} = C_{\mu\alpha\nu\beta} \,n^\alpha\,n^\beta, \label{eq:Eijdef}
\eea
$n^\mu$ being the normal to the constant time slice.
According to~\cite{Clifton:2013jpa}
its evolution is likewise governed by an ODE:
\begin{align}
  \dot E_+ &= -3 \frac{\dot a_\perp}{a_\perp} E_+ \label{eq:cgrt}
\end{align}
so that the evolution of the geometry on the curve is completely decoupled
from its surroundings (in fact, the evolution of every single point on the
curve is decoupled from all the others), and quantities that only depend
on the metric tensor on the curve can be evolved using just the above
system of ODEs (note that we will show in the following sections that (\ref{eq:cgrt})
is missing an essential term which causes this decoupling to fail).

Such a simplified scenario is particularly suitable for use as a numerical testbed,
as one can compare the results of a full three-dimensional numerical evolution
to the functions $a_\parallel$, $a_\perp$ and $E_+$ defined above, and check
to what extent the code reproduces the ODE system. We illustrate the result
of this comparison in the next section.

\section{Numerical Relativity solution of an $S^3$ black-hole lattice spacetime}
\label{sec:num}

\subsection{Methods} \label{sec:nummeth}

We solve the full $3+1$ Einstein equations for an $S^3$ lattice using
Numerical Relativity, allowing us to compute the metric everywhere, not just
on the points of high symmetry. We use the open-source Einstein
Toolkit~\cite{Loffler:2011ay} and Cactus~\cite{Cactuscode:web}
framework.  We compute various lattice-related analysis quantities using a Cactus code generated using
Kranc~\cite{Husa:2004ip,krancweb} and the xAct~\cite{xTensor} tensor-manipulation package.
Analysis of the numerical data was performed
using SimulationTools for Mathematica~\cite{SimulationToolsWeb}.

We focus on the tesseract configuration, in which 8 identical black holes are arranged 
regularly on $S^3$. To simplify the numerical treatment, we
carry out the stereographic projection, introduced in \cite{Bentivegna:2012ei}, from $S^3$ to $R^3$, where one of the black holes,
with bare mass  $m_1 = 4\cal{M}$ is projected into the coordinate origin, another six, each with
bare mass $m_{2-7}=4\sqrt 2 \cal{M}$, are projected to $x^i = (\pm 2, 0,
0)\cal{M}$, $x^i = (0,\pm 2, 0)\cal{M}$, $x^i = (0,0,\pm 2)\cal{M}$, and the eighth
is projected out to infinity (its presence, however, is revealed by an inner trapped surface
at a coordinate radius of about $20\cal{M}$).
Here $\cal{M}$ is a mass parameter, and the seven $m_n$ are the bare masses of the black
holes in the Brill-Lindquist formalism (the asymmetry coming from the stereographic projection,
as their physical masses are all equal to each other).  The 3-metric of the $t=0$ surface is given by:
\be
\gamma_{ij} = \psi^4 \delta_{ij} \label{eq:conformalansatz}
\ee
where $\delta_{ij}$ is the flat metric and $\psi$ is the conformal factor, which takes the form:
\bea
\label{eq:psi}
&&\psi(x,y,z) = 1+ \sum_{n=1}^7 \frac{m_n}{2 r_n} \label{eq:8bhs}
\eea
where $m_n$ is the bare mass of the $n$th black hole and $r_n$ is the distance from it.
The initial data is time-symmetric, so the extrinsic curvature is initially zero;
\bea
K_{ij} = 0. \label{eq:Kij=0}
\eea

We will study one segment of the LDRRS curve corresponding to one of
the edges of the lattice lying on the diagonal between the vertices
$x^i = (2/3, 2/3, 2/3)\cal{M}$ and $x^i = (2, 2, 2)\cal{M}$.  The
$S^3$ angular coordinate $\phi$ along this edge can be related to the
Cartesian coordinates in the stereographic projection via
\bea
\phi &=& \frac{\pi}{2} - \arccos \left ( \frac{3 d^2 - 4}{3 d^2 + 4} \right )  \\
d^2 &=& \frac{4}{3} \left ( \frac{1 + \sin{\phi}}{1 - \sin{\phi}}\right)
\eea
where $d$ is any of the coordinates $x$, $y$ or $z$. $\phi=0$
($d=2/\sqrt{3}$) corresponds to the midpoint of the edge, and most of
our results will be presented using this point as an example.  The
vertices are located at $\phi = \pm \pi/6$.

We refer the reader to~\cite{Bentivegna:2012ei} for a discussion of this initial-data
set, which also clarifies the role of the parameter $\cal{M}$ and the invariance of the
spacetime under a rescaling of $\cal{M}$.

The spatial computational domain is $|x^i| \le
24 \sqrt{3} {\cal M} \approx 41.6 \cal{M}$, and we make use of the reflection symmetry
in the $x$, $y$ and $z$ directions about the coordinate planes through
the origin to restrict the domain to the octant where $x \geq 0$, $y \geq 0$ and $z \geq 0$.  

The Einstein equations are solved using the
McLachlan~\cite{Brown:2008sb} code with fourth order centred finite
differencing for spatial derivatives and fourth order Runge-Kutta for
the time integration. We integrate the Einstein equations in the BSSN
formulation~\cite{Nakamura:1987zz, Shibata:1995we, Baumgarte:1998te},
a variant of the system
(\ref{eq:ADM1})--(\ref{eq:ADM2}).  For more details, we refer the
reader to~\cite{Bentivegna:2012ei}.

In contrast to our previous work in \cite{Bentivegna:2012ei}, we do not use
the standard binary black hole coordinate conditions (1+log slicing and
gamma-driver shift), and then reslice the spacetime in postprocessing so as to
use proper time as a time coordinate. Instead, we have recently found out that one
can evolve this lattice directly in the normal gauge (unit lapse and zero shift),
and still reach a proper time
coordinate of $t \approx 110\cal{M}$ before the metric becomes degenerate in
this coordinate system at the black holes, and further numerical evolution is not possible
\cite{BH}.

In order to assess the effect of the numerical grid spacing on the
results, we compared solutions with different overall grid spacings,
labelled by the number of points, $n$, in one dimension in a certain
coordinate distance.  We report here the results for $n = 32, 40$ and
$48$.  Since the space and time derivatives are computed with fourth
order accuracy in the grid spacing $h \propto 1/n$, and this is
expected to be the dominant source of error, we expect the numerical
error in the solution to scale as $E = O(h^4) = O(n^{-4})$.  Hence,
the errors at the three resolutions should be approximately in the
ratio $1 : 0.41 : 0.20$.

Due to the different length scales in the system, we use mesh
refinement to concentrate the computational grid points in regions
where small length and time scales need to be resolved, and avoid the
prohibitive computational cost of using this same resolution
everywhere.  Mesh refinement is provided by the Carpet
\cite{Schnetter:2003rb} code.  The
coarsest grid (level $L=0$) has a grid spacing of $h_0 =
\mathcal{M}/(\sqrt{3} n)$.  This was chosen so that the midpoint of the edge lies
on a grid point at every resolution to avoid the need to interpolate
data there.  Refined regions are created around each black hole (BH)
and around the edge where we wish to measure quantities accurately.
Several levels of refinement are used, each level having half the grid
spacing and time step of its parent region, resulting in a hierarchy
of nested boxes with all regions on level $L$ completely surrounded by
regions of level $L-1$.  The BH at the origin, the six BHs at $r=2$,
and the edge are on levels $L = 7$, $6$ and $5$ respectively.  The
locations of the boundaries between refinement levels are found to
have a significant effect on high-frequency numerical error measured
on the edge.  This was minimised by ensuring that the boundaries
remain fixed in time and do not intersect the edge.

\subsection{Results and comparison with analytic ODE}
\label{sec:nrcompare}

We now wish to determine whether the Numerical-Relativity (NR)
spacetime satisfies the ODEs derived in~\cite{Clifton:2013jpa} on the
edge of the lattice.  Using the NR spatial metric $\gamma_{ij}$, we compute
$a_\perp$ and $a_\parallel$ as a function of time at the midpoint of
the edge using (\ref{eq:aperp}) and (\ref{eq:apar}).

\begin{figure}
  \centering
  \begin{subfigure}[t]{4.7cm}
    \centering
    \includegraphics{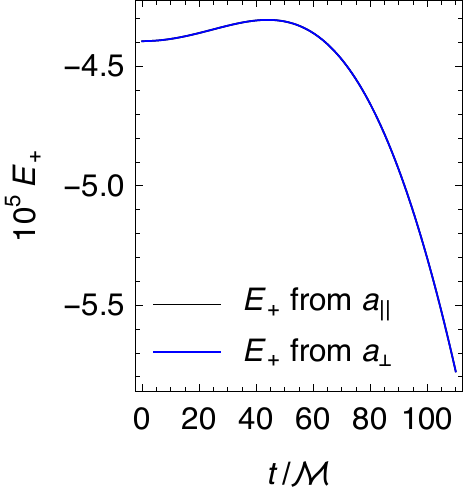}
    \caption{Consistency of $E_+$ computed from the NR $a_\perp$ and $a_\parallel$}
    \label{fig:nrEPlus}
  \end{subfigure}%
  \quad
  \begin{subfigure}[t]{4.4cm}
    \centering
    \includegraphics{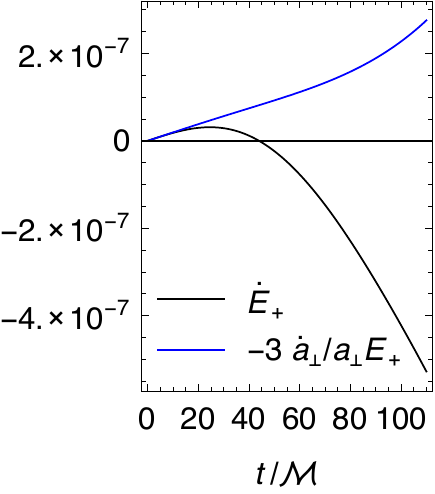}
    \caption{Violation of the ODE $\dot E_+ = -3 \frac{\dot a_\perp}{a_\perp} E_+$ by
      the NR solution}
    \label{fig:nrEPlusDot}
  \end{subfigure}
  \quad
  \begin{subfigure}[t]{5.2cm}
    \includegraphics{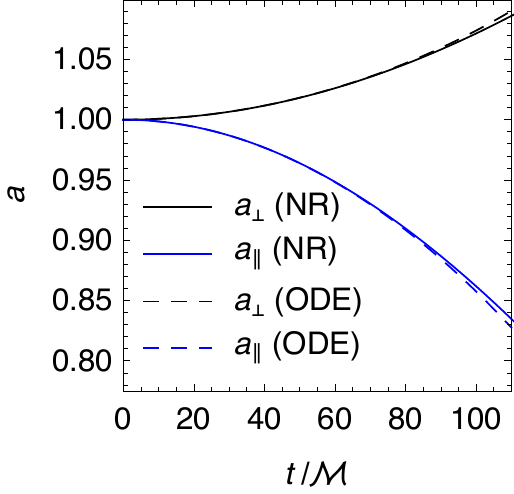}
    \caption{$a_\perp$ and $a_\parallel$ differ by up to 1\% between
      the NR and ODE solutions.  This is much larger than the relative
      numerical error of $10^{-7}$.}
    \label{fig:nradiffs}
  \end{subfigure}
  \caption{Comparison between Numerical Relativity and ODE}
  \label{fig:nroderesults}
\end{figure}

In the~\cite{Clifton:2013jpa} solution, the evolution of $a_\perp$ and
$a_\parallel$ is determined only by $E_+$, and satisfies the system
(\ref{eq:eapardot})--(\ref{eq:aperpdot}). By computing $\ddot
a_\parallel$ and $\ddot a_\perp$ from the NR data, we determine $E_+$
from both equations.  The result is shown in figure \ref{fig:nrEPlus},
and we see that the NR evolution is consistent with
(\ref{eq:aperp}) and (\ref{eq:apar}).

According to~\cite{Clifton:2013jpa}, $E_+$ also evolves according to
the ODE given by (\ref{eq:cgrt})
which involves $a_\perp$ only, so that the system closes and the
solution at the midpoint (as well as of any other point on the edges)
decouples from its surroundings, as there are no spatial derivatives in
the equation. In figure \ref{fig:nrEPlusDot}, we show the NR solution
for $\dot E_+$ and $-3 \frac{\dot a_\perp}{a_\perp} E_+$ which should
agree if the ODE derived in~\cite{Clifton:2013jpa} is correct.  There is a significant
disagreement.  We
have computed the numerical error bars for
figure \ref{fig:nroderesults}, but they are too small to be visible,
indicating that the disagreement is a feature of the continuum Einstein
equations, and not simply due to numerical error.  Figure \ref{fig:nradiffs} shows 
$a_\perp$ and $a_\parallel$ computed from NR and compared with the solution from the ODE.  There
is a disagreement of up to $1\%$ which is not accounted for by the relative
numerical error of $\sim 10^{-7}$.  We thoroughly
investigated all possible sources of error in the NR computation, and
found none that could account for the discrepancy.

For future reference, we define the anomaly $\cal{A}$ as the unknown
additional term in the equation for $\dot E_+$.  Hence, (\ref{eq:cgrt}) from \cite{Clifton:2013jpa} gets modified to:
\begin{align}
  \dot E_+ &= -3 \frac{\dot a_\perp}{a_\perp} E_+ + {\cal A} \label{eq:eplusdot} \, .
\end{align}
In summary, the numerical results suggest that there is a
term, $\cal{A}$, missing in the evolution system
of~\cite{Clifton:2013jpa} which affects $a_\perp$ and $a_\parallel$ at
the level of $1\%$ by $t = 110 {\cal M}$.

\section{The evolution equations on a LDRRS curve}
\label{sec:curl}

We now turn our attention again to the ODE system formed by (\ref{eq:eapardot}), (\ref{eq:aperpdot}) and (\ref{eq:cgrt}), 
and in particular attempt to show that it is the reduction of the system (\ref{eq:ADM1})--(\ref{eq:ADM2}) on a LDRRS curve.

Equations (\ref{eq:ADM1})--(\ref{eq:constraint2}) for $\gamma_{ij}$ and $K_{ij}$ are not closed at a single point because of the presence of the
spatial derivatives of $\gamma_{ij}$ in the 3-dimensional Ricci tensor $R_{ij}$. We will try to close the system by extending it to include $R_{ij}$ as a new evolution variable,
for which we require the time derivative of $R_{ij}$.
 Recall that the time derivative of the 3-dimensional Ricci tensor 
takes the form
\bea
\dot R_{ij} &=& \frac{1}{2}\left( \dot \gamma\UD{k}{i;jk} + \dot \gamma\UD{k}{j;ik} - \dot \gamma\UD{k}{k;ij} - \dot \gamma\DU{ij;k}{k}\right)
\eea
where the indices have been raised by the inverse metric $\gamma^{ij}$, i.e. $\dot \gamma\UD{k}{i} = \dot \gamma_{ji}\,\gamma^{jk}$.
We substitute (\ref{eq:ADM1}) to obtain
\bea
\dot R_{ij} &=& K\UD{k}{k;ij} + K\DU{ij;k}{k}  - K\UD{k}{i;jk} - K\UD{k}{j;ik}.
\eea
We would like to get rid of the second derivatives in the equation above. This obviously requires permuting the indices in the expressions above. Swapping the first or the second pair
of the indices is straightforward, but exchanging two indices between the two pairs is less obvious. We introduce the following notation for the antisymmetrisation in the indices 2 and 3:
\bea
U_{ijkl} = K_{ij;kl} - K_{ik;jl}. \label{eq:Uijkldef}
\eea
After a tedious exercise in index manipulation we arrive at the following expression for the time derivative of the Ricci tensor
\bea
\dot R_{ij} = R\UD{k}{p\left(i\left\|k\right\|\right.}\,K\UD{p}{\left.j\right)} - R\UD{p}{(ij)k}\,K\UD{k}{p} + U\DU{(ij)k}{k}, \label{eq:dotRij3}
\eea
where $\|\cdot\|$ excludes an index from symmetrisation and $R\UD{i}{jkl}$ denotes the Riemann tensor of $\gamma_{ij}$. In three dimensions $R_{ijkl}$ can be expressed entirely via the Ricci tensor and the Ricci scalar~\cite{Wald}:
\bea
R_{ijkl} = R_{jl}\,\gamma_{ik} - R_{il}\,\gamma_{jk} - R_{jk}\,\gamma_{il} + R_{ik}\,\gamma_{jl} + \frac{1}{2}R\left(\gamma_{il}\,\gamma_{jk} - \gamma_{jl}\,\gamma_{ik}\right).
\eea
If we substitute the relation above into (\ref{eq:dotRij3}) we obtain 
\bea
\dot R_{ij} &=& -\frac{3}{2}\left(R\DU{i}{k}\,K_{kj} + R\DU{j}{k}\,K_{ki}\right) + K_{kl}\,R^{kl} \gamma_{ij} \\
  & & + \frac{1}{2}R\,K_{ij} + K\,R_{ij} - \frac{1}{2}R\,K\,\gamma_{ij} + U\DU{(ij)k}{k}, \label{eq:dotRij4}
\eea
i.e. the time derivative of the Ricci tensor expressed directly via $\gamma_{ij}$, $K_{ij}$ and $R_{ij}$ and a single additional term $U\DU{(ij)k}{k}$. The last term, 
i.e. the symmetrized contraction of $U_{ijkl}$, is now the only one involving the spatial derivatives of $K_{ij}$ and $\gamma_{ij}$ and thus not expressible directly via 
$\gamma_{ij}$, $K_{ij}$ and $R_{ij}$. We will introduce a new notation for the symmetrised contraction:
\bea
U_{ij} = U\DU{ijk}{k}. \label{eq:Uijdef}
\eea

\subsection{Properties of $U_{ijkl}$ and relation to the magnetic part of the Weyl tensor}
Before we proceed with the derivation of the reduced Einstein equations, we will discuss some of the properties of $U_{ijkl}$ and elucidate its relation to the magnetic part of the
Weyl tensor. Recall that the Weyl tensor $C_{\mu\nu\alpha\beta}$  in a 3+1
decomposition may be represented by its electric and magnetic parts defined via (\ref{eq:Eijdef}) and
\bea
B_{\mu\nu} = -\frac{1}{2}C_{\mu\alpha\kappa\lambda}\,\eta\UD{\kappa\lambda}{\nu\beta}n^\alpha\,n^\beta \label{eq:Bijdef}
\eea
respectively, 
$n^\mu$ being again the normal to the constant time slice, and
$\eta\UD{\kappa\lambda}{\nu\beta}$ denoting the totally antisymmetric volume form \cite{Nichols:2011pu, ValienteKroon:2005bm}. Both tensors vanish
in the normal direction and can be considered 3-dimensional, spatial objects. 
The magnetic part of the Weyl tensor in the ADM variables can be related to the tensorial curl of the extrinsic curvature:
\bea
B_{ij} = \eta\DU{j}{kl} K_{ik;l} \, . \label{eq:BviaK}
\eea
We can easily prove that it is traceless and symmetric: first, we note that
\bea
B\UD{i}{i} = \eta^{ikl} K_{ik;l} = 0
\eea
because $K_{ik}$ is symmetric with respect to the exchange of the indices. 
The contraction of $B_{ij}$ with another volume form is equal to zero as well:
\bea
B_{ij}\,\eta^{ijp} =&& -\left(\gamma^{ki}\, \gamma^{lp}-\gamma^{kp}\,\gamma^{li}\right)K_{ik;l} = K\UD{pl}{;l} - K^{;p} = 0,
\eea
where the last expression vanishes because of the vector constraint equation (\ref{eq:constraint2}). Note that in three dimensions this implies the vanishing of 
the whole antisymmetric part of $B_{ij}$, so
$B_{ij} = B_{(ij)}$.

The covariant derivative of $B_{ij}$ on the other hand can be related to ${\bf U}$ via
\bea
U_{ijkl} =B_{ip;l}\,\eta\UD{p}{jk}.
\eea
The equation above, contracted with respect to the two last indices, yields
\bea
U_{ij} = -\eta\DU{j}{pl}\,B_{ip;l},\label{eq:UkkviaB}
\eea
which has exactly the same structure as (\ref{eq:BviaK}), i.e. $U_{ij}$ is proportional to the curl of $B_{ij}$.
Now, repeating the reasoning we have used for $B_{ij}$ above we may prove that the trace of (\ref{eq:UkkviaB}) vanishes 
because of the symmetry of $B_{ij}$.\footnote{Note however that $U_{ij}$ does not have to be symmetric in general, unlike $B_{ij}$.}
\bea
U\UD{i}{i} = 0. \label{eq:Utraceless}
\eea
\subsection{The reduced evolution equations}
\label{sec:reducedeqs}

Consider the tangent space at a point along a LDRRS curve $\lambda$. We rewrite (\ref{eq:ADM1})--(\ref{eq:ADM2}) and (\ref{eq:dotRij4}) assuming conditions 
(\ref{eq:GinvX})--(\ref{eq:GinvA}) to hold and parametrizing the metric according to (\ref{eq:apar})--(\ref{eq:aperp}). We first note that the antisymmetric part of $U\DU{ijk}{k}$ must 
vanish because of (\ref{eq:GinvA}).  Since it is
also traceless it must be proportional to $U_{11}$ (see (\ref{eq:GinvS})). We obtain (\ref{eq:eapardot})--(\ref{eq:aperpdot}), where
$E_+ $ is the non-vanishing part of the electric Weyl tensor, but (\ref{eq:cgrt}) now takes the form of
\bea
\dot E_+ &= -3 \frac{\dot a_\perp}{a_\perp} E_+ - \frac{3}{2}U_{11}, \label{eq:neweplusdot}
\eea
with $U_{11} = U\DU{ij}\,\bi{e}_1^i\,\bi{e}_1^j$ (notice that numeric indices always indicate frame components).
In \cite{Clifton:2013jpa}, the authors assume that this term vanishes due to the rotation and reflection invariance.\footnote{It corresponds to the term proportional to $\epsilon^{\gamma\delta\left(\alpha\right.}\bi{e}_\gamma
\left(H \UD{\left.\beta\right)}{\delta}\right)$ in equation (2.15) in the aforementioned paper. If it does not vanish then it appears later in the evolution equation
(4.11).} We will show that this is not 
the case in general.  As a result, we will identify this term with the anomaly $\mathcal{A}$ found numerically in section \ref{sec:nrcompare};
\bea
\mathcal{A} = - \frac{3}{2}U_{11} \, . \label{eq:anomU11}
\eea

First let us consider the magnetic part of the Weyl tensor. Since $B_{ij}$ is composed of rotation-invariant $\eta_{ijk}$ and $K_{ij;k}$ it is rotation-invariant itself. 
Being additionally traceless and symmetric it must be proportional to $B_{11}$ due to (\ref{eq:GinvS}). From (\ref{eq:BviaK}) we obtain
\bea
B_{11} = K_{12;3}-K_{13;2} = 2K_{i[2;3]}\,\bi{e}^i_1 \label{eq:B33viaK}
\eea
Since $\bi{e}_1$ is both rotation- and reflection-invariant, the last expression is the $(2,3)$ component of a rotation- and reflection-invariant rank 2 
antisymmetric tensor, so it must vanish at $\lambda$ 
because of (\ref{eq:GinvA}). 

Now, since $U_{ij}$ is also traceless and is given by a very similar expression (\ref{eq:UkkviaB}) to $B_{ij}$, it would
be tempting to repeat the argument above and conclude that $U_{11}$, together with the whole symmetric part of $U_{ij}$, vanishes too. This would however be 
incorrect due to the following: unlike $K_{ij}$ appearing in (\ref{eq:BviaK}),
$B_{ij}$ in (\ref{eq:UkkviaB}) is \emph{not} reflection-invariant. Note that since its definition (\ref{eq:BviaK}) involves the volume form $\eta_{ijk}$ it changes its sign under 
reflections (\ref{eq:R_m1})--(\ref{eq:R_m3}). Although $U_{11}$ can  be put in a similar form to (\ref{eq:B33viaK}):
\bea
U_{11} = -2B_{i[2;3]}\,\bi{e}^i_1
\eea
we cannot now apply (\ref{eq:GinvA}) because the antisymmetric 2-tensor in question $B_{i[k;l]}\,\bi{e}^i_1$ is \emph{not} reflection-invariant. The symmetry assumptions
put no restrictions on the value of $U_{11}$ and  there no reason whatsoever to assume that $U_{11}$ vanishes
identically along a LDRRS curve.

We can give a simple and instructive analogy from the theory of electromagnetism and Maxwell's equations. Consider a static configuration of electric and magnetic fields in a 
flat space exhibiting a similar rotation and reflection invariance with respect to a chosen axis. If the vector potential $\vec A$ is invariant 
then its curl needs to vanish at the 
symmetry axis and thus the magnetic field $\vec B = \curl \vec A = 0$ along the axis due to the reflection symmetry, just like in the case of the magnetic Weyl tensor. 
But the curl of $\vec B$ \emph{does not} need to vanish at the axis.
Indeed, it 
is easy to create a configuration of the electromagnetic field in which there is a non-vanishing current flowing along the axis and thus $\vec j = \curl \vec B \neq 0$. 
This is due to the fact that $\vec B$, as a curl of a vector, is a pseudovector field, while $\vec j$, which is a \emph{curl of a curl} of a vector is again a regular vector field.
The former must vanish because of the reflection symmetry, but the latter not.

\subsection{$U_{11}$ and its time derivatives for the 8-black-hole initial data}

The initial data described in section \ref{sec:nummeth} is time-symmetric, so the solution in normal coordinates satisfies $\gamma_{ij}(t) = \gamma_{ij}(-t)$. It follows that
the odd time derivatives of the metric $\frac{\partial^{2N+1}}{\partial t^{2N+1}} \gamma_{ij}$ and of the Christoffel symbols $\frac{\partial^{2N+1}}{\partial t^{2N+1}}\Gamma\UD{i}{jk}$,
$N = 0,1,2\dots$,
vanish at $t=0$ identically. So does the extrinsic curvature together with its even time derivatives $\frac{\partial^{2N}}{\partial t^{2N}} K_{ij}$. 
From (\ref{eq:Uijkldef}) and (\ref{eq:Uijdef}) we see that the same must hold for $U_{ij}$, i.e.
\bea
 \frac{\partial^{2N}}{\partial t^{2N}} U_{ij} = 0 \qquad\textrm{at }t=0
\eea
and in particular $U_{ij}$ and $\ddot U_{ij}$ vanish initially. 
Direct computation reveals that for the initial data (\ref{eq:Kij=0}) and (\ref{eq:conformalansatz}) the first
 derivative $\dot U_{ij} = 0$ along a LDRRS curve vanishes as well. 
The first non-vanishing time derivative turns out to be the third one. We have evaluated it as a combination of the partial derivatives of the conformal factor $\psi$. Since
the expression involves up to 4th covariant derivatives of the Ricci tensor of $\gamma_{ij}$, the tensor manipulations and algebraic reduction were performed using Mathematica. 
The final result, considered along the LDRRS curve and after simplifications due to the symmetry, reads
\bea
U_{11}^{(3)}\big|_{t=0} &&=2\psi^{-18} \left(-\left(\psi ^{(0,0,6)}+3 \psi ^{(0,2,4)}+3 \psi ^{(0,4,2)}+\psi ^{(0,6,0)}\nonumber\right.\right.\\
&&\left.\left.
+2 \left(\psi ^{(2,0,4)}+2 \psi ^{(2,2,2)}+\psi ^{(2,4,0)}\right)-\psi ^{(6,0,0)}\right) \psi ^5\right.\nonumber\\
&&\left.
+\left(-8 (\psi ^{(3,0,0)})^2+\psi ^{(2,0,0)} \left(35 \psi ^{(4,0,0)}-29 \left(\psi ^{(0,0,4)}+2 \psi ^{(0,2,2)}+\psi^{(0,4,0)}\right)\right)\right.\right.\nonumber\\
&&\left.\left.
+2 \left(24 (\psi ^{(0,0,3)})^2+44 \psi ^{(0,2,1)} \psi ^{(0,0,3)}+28 (\psi ^{(0,1,2)})^2+28 (\psi ^{(0,2,1)})^2+24(\psi^{(0,3,0)})^2\right.\right.\right.\nonumber\\
&&\left.\left.\left.
+44 \psi ^{(0,1,2)} \psi ^{(0,3,0)} +6\psi ^{(0,1,1)} \left(\psi ^{(0,1,3)}+\psi ^{(0,3,1)}\right)+7 \psi ^{(1,0,1)} \left(\psi ^{(1,0,3)}+\psi^{(1,2,1)}\right)
\right.\right.\right.\nonumber\\
&&\left.\left.\left.+7 \psi ^{(1,1,0)} 
\left(\psi ^{(1,1,2)}+\psi ^{(1,3,0)}\right)+2 \psi ^{(1,0,0)} \left(\psi ^{(1,0,4)}+2 \psi ^{(1,2,2)}+\psi ^{(1,4,0)}-\psi ^{(5,0,0)}\right)\right)\right) \psi ^4\right.\nonumber\\
&&\left.
-2 \left(18 (\psi ^{(2,0,0)})^3-24 \psi ^{(1,0,0)} \psi ^{(3,0,0)} \psi ^{(2,0,0)}+2 \psi ^{(0,0,4)} (\psi ^{(0,1,0)})^2+11 (\psi ^{(0,0,1)})^2 \psi ^{(0,0,4)}\right.\right.\nonumber\\
&&\left.\left.
+18 \psi ^{(0,0,1)} \psi ^{(0,1,0)} \psi ^{(0,1,3)}+13 (\psi ^{(0,0,1)})^2 \psi ^{(0,2,2)}+13 (\psi ^{(0,1,0)})^2 \psi ^{(0,2,2)}
\right.\right.\nonumber\\
&&\left.\left.
+18 \psi ^{(0,0,1)} \psi ^{(0,1,0)} \psi ^{(0,3,1)}+
2 (\psi ^{(0,0,1)})^2 \psi ^{(0,4,0)}+11 (\psi ^{(0,1,0)})^2 \psi ^{(0,4,0)}+ \right.\right.\nonumber
\eea
\bea
&&\left.\left.
+7 \psi ^{(1,0,0)} \left(7 \left(\psi ^{(0,0,3)}+\psi ^{(0,2,1)}\right) \psi ^{(1,0,1)}+
7 \left(\psi ^{(0,1,2)}+\psi ^{(0,3,0)}\right) \psi ^{(1,1,0)}\right.\right.\right.\nonumber\\
&&\left.\left.\left.+3 \psi ^{(0,0,1)} \left(\psi ^{(1,0,3)}+\psi ^{(1,2,1)}\right)+3 \psi ^{(0,1,0)} \left(\psi ^{(1,1,2)}+
\psi ^{(1,3,0)}\right)\right)-8 (\psi ^{(0,0,1)})^2 \psi ^{(4,0,0)}\right.\right.\nonumber\\
&&\left.\left.
-8 (\psi ^{(0,1,0)})^2 \psi ^{(4,0,0)}+(\psi ^{(1,0,0)})^2 \left(26 \psi ^{(4,0,0)}-20 \left(\psi ^{(0,0,4)}+
2 \psi ^{(0,2,2)}+\psi ^{(0,4,0)}\right)\right)\right) \psi ^3\right.\nonumber\\
&&\left.
-6 \left(3 \left((\psi ^{(0,0,1)})^2+(\psi ^{(0,1,0)})^2\right) (\psi ^{(2,0,0)})^2-\psi ^{(1,0,0)}
\left(49 \left(\psi ^{(0,0,1)} \left(\psi ^{(0,0,3)}+\psi ^{(0,2,1)}\right)\right.\right.\right.\right.\nonumber\\
&&\left.\left.\left.\left.
+\psi ^{(0,1,0)} \left(\psi ^{(0,1,2)}+\psi ^{(0,3,0)}\right)\right) \psi ^{(1,0,0)}-8 
\left((\psi ^{(0,0,1)})^2+(\psi ^{(0,1,0)})^2\right) \psi ^{(3,0,0)}\right)\right) \psi ^2\right.\nonumber\\
&&\left.+288 \left((\psi ^{(0,0,1)})^2+(\psi ^{(0,1,0)})^2\right) (\psi ^{(1,0,0)})^2 
\psi ^{(2,0,0)} \psi\right.\nonumber\\
&&\left. -288 \left((\psi ^{(0,0,1)})^2+(\psi ^{(0,1,0)})^2\right) (\psi ^{(1,0,0)})^4\right),\label{eq:d3u33analytic}
\eea
where we have assumed above that the first coordinate $x^1$ is aligned along the curve and $x^2, x^3$ are transversal. We 
have introduced here a short hand notation for the partial derivatives in the form of $\psi^{(p,q,r)} = \frac{\partial^p}{\partial (x^1)^p}
\frac{\partial^q}{\partial (x^2)^q}\frac{\partial^r}{\partial (x^3)^r}\psi$. 
Substituting the conformal factor $\psi$ from equation (\ref{eq:psi}),
we obtain that the numerical value of $U^{(3)}_{11}(0)$ at the midpoint of the edge is $4.3\times 10^{-12}$.

\subsection{Effect of $U_{11}$ on the metric}
\label{sec:u11metrictaylor}

The addition of the term $-3/2 U_{11}$ to the ODE clearly affects the
evolution of $a_{\parallel}$ and $a_{\perp}$.  We can estimate the
effect by making a Taylor expansion of $a_{\parallel}(t)$ and
$a_{\perp}(t)$ about $t=0$ and using the evolution equations
(\ref{eq:aperpdot}), (\ref{eq:eapardot}) and (\ref{eq:neweplusdot}) to
evaluate the Taylor coefficients at $t=0$.  We find that the effect of
$U_{11}$ appears first in the $O(t^6)$ term.  Using an overbar to
represent the solution using the original ODE (\ref{eq:cgrt}),
i.e.~without the $U_{11}$ term, we find
\begin{alignat}{2}
\Delta a_{\parallel} &=  a_{\parallel} - \bar a_{\parallel} &&
= - \frac{a_{\parallel}(0)}{720} U^{(3)}_{11}(0) t^6 + O(t^8) \, , \\
\Delta a_{\perp}    &=  a_{\perp} - \bar a_{\perp} &&
= \frac{a_{\perp}(0) }{1440} U^{(3)}_{11}(0) t^6 + O(t^8) \, .  \\
\end{alignat}
where we have used the fact that $a_{\perp}$, $a_{\parallel}$ and
$E_+$ have the same value at $t=0$ independent of the appearance of
$U_{11}$ in the ODE.  

At the midpoint of the edge, at $t = 110 \mathcal{M}$, we find a
relative error $\Delta a_{\parallel}/a_{\parallel}$ of about $1\%$
compatible with the NR results in figure \ref{fig:nradiffs}, and a
relative error of $100\%$ by $t = 235 \mathcal{M}$.  We conclude that
the leading order contribution to $U_{11}$ leads to a complete
breakdown of the original ODE solution by this time, though we cannot
determine whether higher order corrections are important here.

\section{Numerical Relativity calculation of $U_{11}$}
\label{sec:numu33}

\subsection{Consistency between NR and new analytical results}

In the previous section, we identified a term, $-3/2 U_{11}$, in the
evolution equation for $\dot E_+$ which was assumed in
\cite{Clifton:2013jpa} to vanish, but for which we find a nonvanishing
third time derivative.  We now aim to verify that the NR solution
satisfies the new evolution equation (\ref{eq:neweplusdot}), and that
the numerically-non-zero anomaly $\mathcal{A}$ is indeed related to
$U_{11}$ by (\ref{eq:anomU11}).  $U_{11}$ is computed in NR
from covariant derivatives of the extrinsic curvature,
\begin{align}
  U_{11}   &= ({K_{ij;k}}^{k} - {K_{ik;j}}^k)\, \bi{e}_1^i\, \bi{e}_1^j \, . \label{eq:u33}
\end{align}
whereas ${\cal A}$ is defined via (\ref{eq:eplusdot}).

\begin{figure}
  \centering
  \begin{subfigure}[t]{5.9cm}
  \centering
    \includegraphics{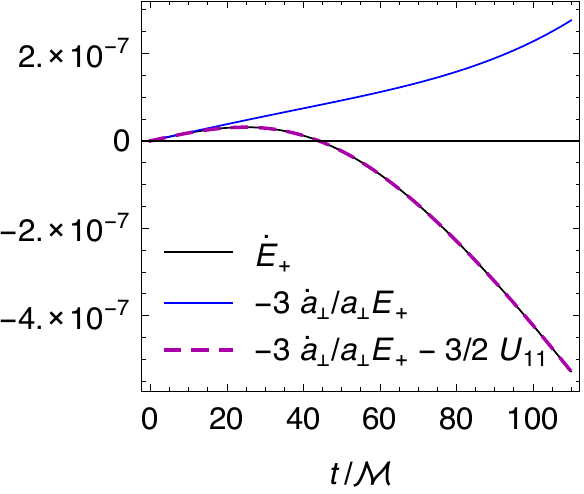}
    \caption{Comparison of $\dot E_+$ with the original and new RHSs}
    \label{fig:nreplusdotfixed}
  \end{subfigure}%
  \quad
  \begin{subfigure}[t]{4.3cm}
  \centering
    \includegraphics{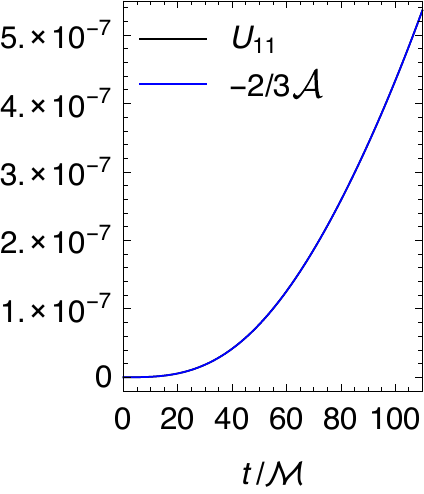}
    \caption{$U_{11}$ and $-2/3 {\cal A}$ computed from NR}
    \label{fig:nrU33A2}
  \end{subfigure}%
  \quad
  \begin{subfigure}[t]{4.3cm}
    \includegraphics{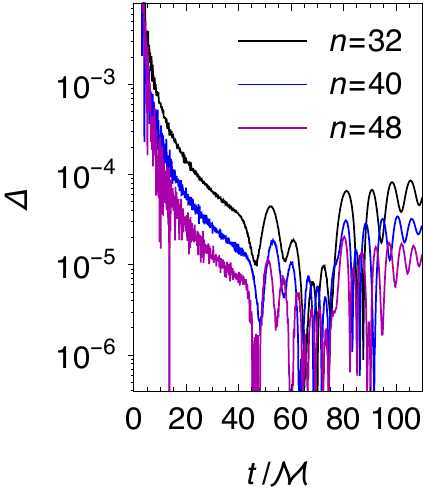}
    \caption{$\Delta \equiv 1-(-2/3\mathcal{A})/U_{11}$ computed from NR
      at several resolutions.}
    \label{fig:nrA3Conv}
  \end{subfigure}
\caption{NR solution demonstrating consistency with the new analytic results}
\label{fig:nrresults}
\end{figure}

Figure \ref{fig:nreplusdotfixed} shows a comparison between $\dot E_+$
and the RHS of the original and new evolution equations.  We see that
the addition of the term $-3/2 U_{11}$ is necessary for agreement.
In figure \ref{fig:nrU33A2}, we see that $U_{11}$ and $-2/3
\mathcal{A}$ are found to be indistinguishable, and figure
\ref{fig:nrA3Conv} shows that their relative difference, $\Delta
\equiv 1-(-2/3\mathcal{A})/U_{11}$, converges to zero as the numerical
resolution $n$ is increased.  The convergence is 4th order, as
expected from the finite differencing order of the code.  $\Delta$
exhibits high-frequency noise for $t<40\mathcal{M}$ which we attribute
to error coming from
the finite precision with which floating point numbers are represented in
the code\footnote{${\cal A}$ depends on the third time
  derivative of $a_\perp$, and an initial relative roundoff error of
  $\epsilon \sim 10^{-15}$ with frequency $\omega \sim \pi/\Delta t
  \sim 80$, for $\Delta t$ the time spacing of output data points,
  will be amplified by a factor of $a_\perp/\dddot a_\perp
  \omega^3$ when taking a third derivative, which leads to a relative
  error in $\dddot a_\perp$ comparable with that observed for the
  measured values of $a_\perp \sim 10^2$ and $\dddot a_\perp \sim
  10^{-7}$.}.
We have partially filtered the high frequency noise from the data in
figure \ref{fig:nrA3Conv} to make the convergence more apparent.  For
$t>40\mathcal{M}$, there are lower-frequency oscillations in the error
which we attribute to numerical reflections from mesh refinement
boundaries.

For $t > 20 {\cal M}$, at the highest resolution, we see that
$|\Delta| < 3 \times 10^{-5}$.  Hence
\begin{align}
-\frac{2}{3} {\cal A} = 1.00000(3) \, U_{11}
\end{align}
in agreement with the analytic derivation in
section \ref{sec:reducedeqs}. For $t < 20 \mathcal{M}$, the ratio is still
consistent with $-2/3$,
but the relative error is larger since $U_{11}$ itself is small.

We therefore see that the anomaly originally
measured in the comparison of the 3+1 Numerical Relativity results and the
ODE system presented in~\cite{Clifton:2013jpa} was due to the term $U_{11}$ derived in 
the previous section, but taken to vanish in the original derivation.

\subsection{Computation of $U_{11}$ and fitting formula}

\begin{figure}
  \centering
  \includegraphics{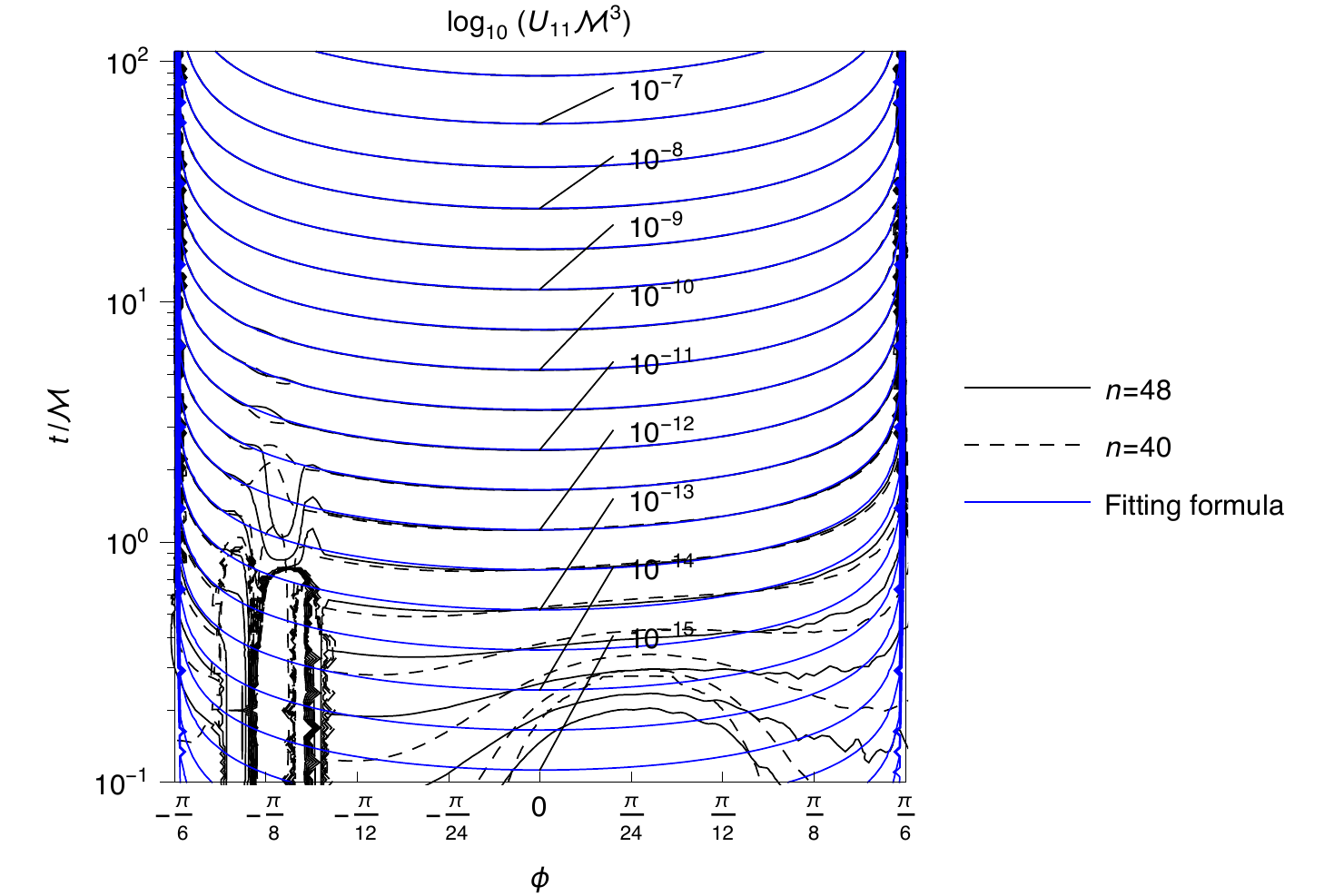}
  \caption{Contours of $\log_{10} U_{11}$ computed using NR at two
    different resolutions (solid and dashed black contours) and from
    a simple polynomial formula fitted to the NR data (blue
    contours) as a function of $t$ (time) and $\phi$ (coordinate
    along the edge). The fit is performed in the region $1{\cal M} \le t
    \le 110{\cal M}$.}
  \label{fig:nrU33contours}
\end{figure}

We now present the NR computation of $U_{11}$ on the edge, and give a
simple fitting formula for it that could be used along with
(\ref{eq:neweplusdot}) to solve the system via an ODE.

Figure \ref{fig:nrU33contours} is a contour plot of $\log_{10} (U_{11} {\cal M}^3)$ as a function of $t$ and $\phi$, the proper time and the $S^3$ angular
coordinate along the edge, respectively.
The black solid and dashed lines represent contours of $U_{11}$ computed at
resolutions $n=48$ and $n=40$ respectively.  For $t \ge 10\cal{M}$ the two
resolutions are indistinguishable, indicating that the numerical error
is small in comparison with $U_{11}$.  For $t < 10\cal{M}$, there are
regions, notably around $\phi \sim -\pi/8$, where the numerical error
dominates over $U_{11}$, which at these early times is $O(10^{-15})$.
The NR data satisfies $U_{11}(0,\phi) = U_{11}(t,\pm \pi/6) = 0$ on the
initial slice and the vertices as expected, since $U_{11} = 0$ there by symmetry.
Note that the NR computation was performed in Cartesian coordinates
$(t, x, y, z)$ and has been transformed to $(t, \phi = \frac{\pi
}{2}-\cos ^{-1}\left(\frac{3 x^2-4}{3 x^2+4}\right))$ for plotting.  While the
spacetime is symmetric about $\phi=0$ in $(t,\phi)$, this is not the
case in $(t,x,y,z)$, hence the continuum solution is expected to show
this symmetry in $\phi$, but the numerical error is not.  This is
reflected in Figure \ref{fig:nrU33contours}.

Since $U_{11}$ is only available numerically, and appears to have a
simple form in the regions in which it is well resolved, we provide a
simple formula based on low-order polynomials in $t$ and $\phi$
obtained via a least-squares fit to the NR data in the region $1 \mathcal{M} \le t
\le 110 \mathcal{M}$, $-\pi/6 \le \phi \le \pi/6$, corresponding to the edge of
the lattice.  The fitting formula is
\begin{equation}
U_{11} = \sum_{p,q} c_{pq} (t/{\cal M})^p \phi^q {\cal M}^{-3} \quad p=3,5,7 \quad q=0,2,4,6
\label{eq:nrfitformula}
\end{equation}
with the coefficients $c_{pq}$ of $(t/{\cal M})^p$ and $\phi^q$ given in Table
\ref{tbl:U33coeffs}. The error estimate in the last digit in
parentheses is an indication of numerical truncation error.
The contours of the fitting formula are shown in
blue in Figure \ref{fig:nrU33contours}.

\begin{table}
\begin{tabular}{>{$}c<{$} | d{7} @{} >{$}r<{$} d{6} @{} >{$}r<{$} d{5} @{} >{$}r<{$} d{5} @{} >{$}r<{$} }
  \text{} & 1 & & \phi ^2 & & \phi ^4 & & \phi ^6 & \\
 \hline
(t/{\cal M})^3 & 7.040(1) & \times 10^{-13} & -3.558(6) & \times 10^{-12} & 4.23(4) & \times 10^{-12} & -2.30(7) & \times 10^{-12} \\
(t/{\cal M})^5 & -3.810(2) & \times 10^{-17} & 2.361(9) & \times 10^{-16} & -4.37(6) & \times 10^{-16} & 3.1(1) & \times 10^{-16} \\
(t/{\cal M})^7 & 1.0965(9) & \times 10^{-21} & -6.85(3) & \times 10^{-21} & 1.30(2) & \times 10^{-20} & -9.8(3) & \times 10^{-21} \\
\end{tabular}

\caption{Coefficients of $(t/{\cal M})^p$ and $\phi^q$ in the fitting formula for
  $U_{11}$ determined from NR}
\label{tbl:U33coeffs}
\end{table}
The region $1{\cal M} \le t \le 10{\cal M}$ contains localised regions of high
relative numerical error, but the small number of degrees of freedom
in the fitting formula means that the fit is insensitive to these
localised regions.
The region $t \le 1\cal{M}$, in which the NR error
dominates, is outside the fit region, and hence the fitting formula is
an extrapolation in this region.
For $t > 10 {\cal M}, |\phi| < \pi/8$, i.e.~the regions where $U_{11}$
is not close to zero, this fitting function approximates the NR result
to within $\pm 1\%$.  In the regions $t < 10 {\cal M}$ and $|\phi| >
\pi/8$, the absolute agreement is within $10^{-12} \mathcal{M}^{-3}$.

For $t \ge 10\cal{M}$, the NR and fitting-formula curves are visually
indistinguishable.

\subsection{Computation of $U_{11}^{(3)}$}

We now wish to compute the third time derivative of $U_{11}$ at
$\phi=0$ from the NR data and compare with the analytic result
obtained in (\ref{eq:d3u33analytic}).  We cannot directly
finite-difference the NR data near $t = 0$ because, as can be seen in
figure \ref{fig:nrU33contours}, it is contaminated by numerical error.
Instead, we compute the derivative by analytically differentiating the
fitting formula.  The fitting effectively averages out the very small
numerical errors near $t = 0$ and uses information from $t > 0$, where
the errors are less significant, to obtain information about the
derivative at $t=0$.

The fitting formula (\ref{eq:nrfitformula}) contains only a finite
number of terms, so the coefficients cannot be directly identified
with the coefficients in a Taylor series, and hence with the
derivatives of $U_{11}$.  However, as the number of terms in the
fitting formula is increased, we expect the coefficients to approach
the Taylor coefficients.  We find that as both $p_\mathrm{max}$ and
$q_\mathrm{max}$ are increased, $c_{30}$ appears to converge
exponentially towards a limiting value.  Taking this to be the Taylor
coefficient, we obtain an NR estimate for $U^{(3)}_{11}$ which can be
directly compared with the analytic value obtained from
(\ref{eq:d3u33analytic}):
\begin{align}
  \frac{\partial^3 U_{11}}{\partial t^3} \bigg |_{t=0,\phi=0} =
  \begin{cases}
    4.3015(4) \times 10^{-12} {\cal M}^{-6} & \text{Numerical} \\
    4.30113   \times 10^{-12} {\cal M}^{-6} & \text{Analytic} \, .
  \end{cases}
\end{align}
The NR error estimate in parentheses includes the effect of both
numerical truncation error and of fitting using a finite number of
terms, and we see that the NR derivative matches the analytical
calculation within NR errors.  We therefore have a high degree of
confidence that the numerical solution and our understanding of the
analytical system are correct.

\subsection{Effect of $U_{11}$}

\begin{figure}
  \centering
  \includegraphics{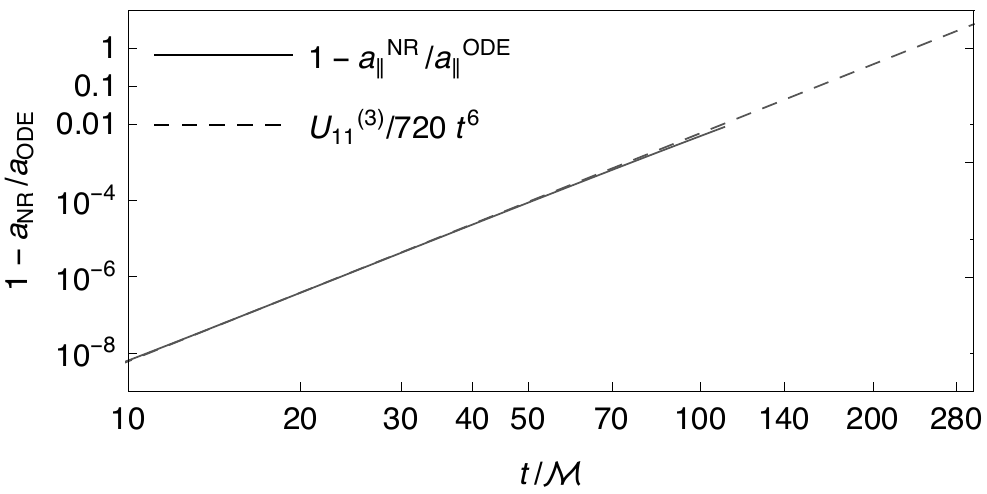}
  \caption{Comparison between the difference between the NR and ODE
    solutions for $a_{\parallel}$, and the leading order contribution
    computed from a Taylor expansion.}
  \label{fig:nrtaylorplot}
\end{figure}

In figure \ref{fig:nrtaylorplot}, we show the relative difference
between $a_{\parallel}$ computed from NR and from the original ODE,
and compare it with the leading order analytic contribution computed
in section \ref{sec:u11metrictaylor} from the Taylor series.  We see
that the relative difference is dominated by the leading order term
for as long as the NR computation lasts.  We do not know whether this
will continue past $t = 110 \mathcal{M}$.

\section{Conclusions}
\label{sec:concl}

We have solved the full Einstein equations for an $S^3$ 8-black-hole
lattice spacetime using Numerical-Relativity and found that the
solution along certain LDRRS curves does not agree with the ODEs
previously derived for this system.

We therefore analysed the behaviour of LDRRS curves in vacuum
spacetimes, and in particular showed that the evolution of certain symmetric subsets
does not decouple from the surrounding spacetime, following a system of pure ODEs,
as had been previously claimed. Instead, the variables $a_\parallel$, $a_\perp$ and 
$E_+$, which capture all the metric and extrinsic-curvature degrees of freedom
not suppressed by the symmetries, follow a system of ODEs with a source term
$U_{11}$, which itself depends on the spatial derivatives of the extrinsic curvature
and can only, to our knowledge, be computed via Numerical Relativity.

We have then computed this term using Numerical Relativity, both as an 
anomaly in the original ODE system, and via its expression in terms of the
derivatives of $K_{ij}$. These agree to within the relative numerical error
of $3 \times 10^{-5}$,
strengthening our confidence in both the analytical study performed in sections \ref{sec:lrs}
and~\ref{sec:curl} and the numerical infrastructure used in sections \ref{sec:num} 
and~\ref{sec:numu33}, as 
well as in~\cite{Bentivegna:2012ei}.

Note that it is still possible to use the ODE system presented in~\cite{Clifton:2013jpa},
as long as one knows the source term independently. To this end, we provide
a polynomial function, fitted from the NR data, in both proper time $t$ and edge coordinate $\phi$, 
which can be used up until $t \sim 110 \cal{M}$. We have also computed
the third time derivative of $U_{11}$ on the midpoint of the edge at $t=0$ (which is the lowest
non-zero time derivative of $U_{11}$ initially), and compared it with an analytical
derivation of the same quantity in the ADM formalism. These also agree within the
relative numerical error of $10^{-4}$.

We conclude by remarking that the evolution of the edges of $S^3$ black-hole
lattices is not symmetric enough for a reduced-dimension calculation\footnote{Note 
that the vertices, on the other hand, possess enough symmetries
for a full decoupling from the neighbouring points, leading to the solution
$E_+=0$ and the 3-metric being constant.}, and can, to our knowledge, only be
performed via Numerical Relativity. In particular, the solution to the ODE
system presented in~\cite{Clifton:2013jpa} can be treated as an approximation,
valid at early times, which may or may not be close to the true solution at
late times. We have measured the error associated with such an
approximation by direct comparison with a 3+1 integration of the
Einstein equations, and find it to grow to $\sim 1\%$ in the components of
the spatial metric for $t \lesssim 110{\cal M}$.  We have shown analytically
that the leading
order effect of the $U_{11}$ term is $O(t^6)$ in the metric, and this is observed
to a very good approximation in the NR results for 
$t < 110 \mathcal{M}$.  The duration of the NR
computations presented here is limited by the use of the normal gauge,
and at the present time, we have no way of assessing the error
resulting from neglecting the $U_{11}$ term at late times far from the
time-symmetric hypersurface.  We observe that if the $t^6$ growth were
to continue, the metric would have $100\%$ error by $t \sim 235
\mathcal{M}$.
At late times, the results obtained from the system derived
in~\cite{Clifton:2013jpa} may well be qualitatively different to those
that would be obtained by an evolution using the full, corrected,
system.

\section*{Acknowledgements}
MK and EB would like to thank the Max Planck Institute for Gravitational Physics (Albert Einstein Institute) in Potsdam for hospitality. The work was supported by the project \emph{
``The role of small-scale inhomogeneities in general relativity and cosmology''} (HOMING PLUS/2012-5/4), realized within the Homing Plus programme of
Foundation for Polish Science, co-financed by the European Union from the Regional Development Fund, and by the project 
``\emph{Digitizing the universe: precision modelling for precision cosmology}'', funded by the Italian Ministry of Education, University and Research (MIUR).  
The Numerical Relativity computations were performed on the Datura cluster at the AEI. Code for solving the ODEs was written in collaboration with Miko\l{}aj Bi\'nkowski. We thank Timothy Clifton for reading a draft of this manuscript and pointing out several typographical errors.
Any remaining mistakes and omissions remain the responsibility of the authors.

\appendix

\bibliographystyle{jcap/JHEP}
\bibliography{refs}

\end{document}